# Agent-Based µ-Tools Integrated into a Co-Design Platform

Alain-Jérôme Fougères[1]

[1] **University of Technology of Belfort-Montbéliard**
90010 Belfort, France
*alain-jerome.fougeres@utbm.fr*

**Abstract**
In this paper we present successively the proposition and the design of: 1) µ-tools adapted to collaborative activity of design, and 2) a multi-agent platform adapted to innovative and distributed design of products or services. This platform called *PLACID* (innovating and distributed design platform) must support applications of assistance to actors implies in a design process that we have called µ-tools. µ-tools are developed with an aim of bringing assistance to Co-design. The use of the paradigm agent as well relates to the modeling and the development of various layers of the platform, that those of the human-computer interfaces. With these objectives, constraints are added to facilitate the integration of new co-operative tools.
***Keywords:*** *multi-agent system, co-operative agents, co-design, distributed design, micro-tools, development process.*

## 1. Introduction

The objectives of this paper are to present: first, the design of a platform adapted to the innovative and distributed design, and secondly, assistance applications to actors of design process that the platform supports. These applications are called micro-tools (µ-tools) [23, 15]. The concept of µ-tools consists of software applications which are light, easy to use, integrated in a shared environment, and connected between them using a database. This Platform (*PLACID*: platform for innovating and distributed design, *Plate-forme Logiciel d'Aide à la Conception Innovante et Distribuée* in French) is developed with an aim of bringing an assistance to the work of co-design, guided or not by complex processes (like workflow) for their capacity to manage flows of co-operative work (control and execution of co-operative processes). In addition to these objectives, strong constraints of flexibility and adaptability are added, to facilitate the integration of new co-operative tools.

We remember that, in a general way, co-operative activities integrated in virtual spaces of design require tools for:
- Interpersonal or group communication (synchronous and/or asynchronous communication tools).
- Organization and cohesion of the groups and the activities (coordination tools).
- Distribution and division of information, applications and resources (distribution and sharing tools).
- Space-time definition of co-operation: space distance between the members of a team (in a real or virtual room), and temporal distance in the interaction (sequentially or parallelism of tasks).

The co-operative design platform that we present in this paper is based on a software agent approach – approach well adapted for distribution of components. The principal characteristics of agents (autonomy, adaptability, co-operation and communication) make it possible, first, to effectively manage distributed, heterogeneous and autonomous components, and secondly, to facilitate exchanges of information and resource sharing between the components (interaction, communication and co-operation). These agents are of type: application (µ-tools or other tools of assistance to Co-design), coordinator/mediator, system and interface. Agents based system must manage organization and control of the community of agents. The effective use of the co-operative design platform (via an interface itself agent-based) is done in a context of strong and multiple interactions, multi-users and multi-modalities.

The µ-tools supported by this platform will not be necessarily integrated in a preset process of design. Their use can be specific, bringing a quite precise service in a phase of design. In all cases, each µ-tool will be connected to the multi-agent system (MAS) by the intermediary of a host agent. This one will be used as interface of communication (inputs/outputs) between µ-tools and co-operative information system.

This article is structured as follows: in section 2 we present the concepts implied in Co-design activities. The following section describes capabilities of agent to communicate and interact. Sections 4 and 5 successively present objectives of *PLACID* platform and design of the first set of µ-tools intended to validate this platform. Finally, in section 6, we discuss the prospects of our work.

## 2. µ-tools for Collaborative Work

2.1 Context of Co-design

The development of data-processing technologies, the democratization of the Internet, the use of the new resources on Internet gave rise to new working methods. We speak of course, of the Computer Supported Cooperative Work (CSCW) [5, 11, 21]. One of the major topics in the field of the CSCW is the development of groupwares. By definition a groupware is software which assists a user group for realization of a joint project. Group members collaborate remotely, either at the same moment (synchronous activity), or at different times (asynchronous activity). The fields of application are very numerous: products design, teaching, trade or games. Groupwares must make it possible to several users to collaborate in explicit shared spaces. The concepts to be considered are as follows [5, 6]:

- *Time and space*: to bring together several distant people geographically (office in proximity or distant) and/or not working at the same time (different rhythms, incompatibility of the timetables...).
- *Modes of co-operation*: asynchronous co-operation (autonomous working method), co-operation in session (the objective being to decrease the times of interaction between the various actors of a project), co-operation in meeting (the roles of actors are defined and each one takes part in its turn), close co-operation (increase in co-production).
- *Flexibility in heterogeneous fields*: interactions, distribution of data, resource sharing, access control, representation of information, planning for tasks execution.

Activities of co-operative and distributed design are exchanges, division and co-operation between participants. It is usual to present co-operative information systems like being able to answer to the different needs of co-operation:
- Facilitating the resource sharing.
- Assisting the coordination.
- Improving the communication of group.
- Supporting the individual motivation.
- Supporting the development of organization.

Activities related to collaborative work are mostly exchanges (language acts, transactions...), sharing and cooperation among participants. Then, to memorize easily the functions of groupware applications, it is convenient to take into account five main functions, namely: 1) *Communication* between participants of the community; 2) *Coordination* between them – this function can be insured by a centralized way or not; 3) *Co-memorization*, which means community's memory construction – it may be a project traceability support, like the set of sheets of paper produced during the course of action; 4) *Co-production* of shared resources, like proxemic space of cooperation, shared objects; and 5) Control of processes, control data or files circulation – it may be workflow, which is a convenient software application if photographed evaluators The following figure (see Figure 1) schematizes relations between these basic functionalities of groupware. We have called this functional model: *5Co* [15]. This model supplements the model *3C* (communication, co-operation, coordination) [6], defining spaces necessary to the artifacts of collaboration.

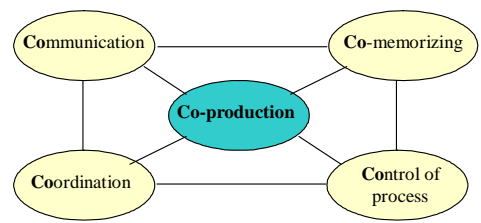

Fig. 1 Basic functions for Co-design – The *5Co*

How to design software applications that should achieved the *5-Co* model, and that both in a technocentric and in an anthropocentric point of view? CSCW approach can be complemented by a more "microscopic" one, focused on course of action, within many different operations are realized, at a very short term, and by cyclical and opportunistic way. To aid these tasks, a new software concept can be defined. It is called "micro-tool" (µ-tool).

2.2 Concept of µ-Tool

The concept of µ-tool [23, 15] is opposed to the current tendency tools of design, which are often heavy, prescriptive, and not used. Ideally, these tools must be (see Figure 2):
- Easy to learn (a few minutes) and easy to use.
- Simple (even if they are developed on the basis of an elaborate theory).
- Easily programmable, easily modifiable by designers or by users themselves, usable in an opportunist way.
- Autonomous, but also reactive when they are defined for co-operative processes (*CMT*: Co-operative Micro-Tools). Those are distributed between actors who act according to their skills – this corresponds to the needs for concurrent engineering.

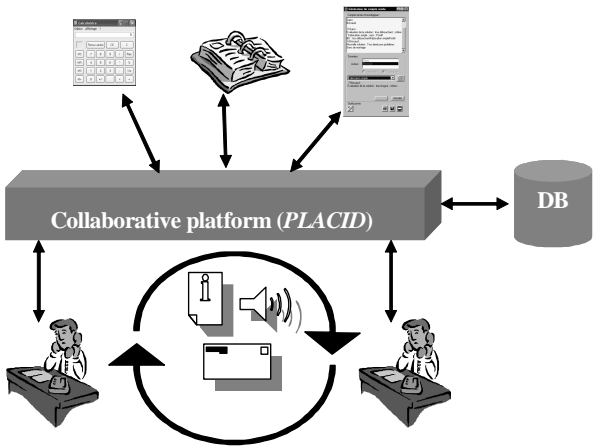

Fig. 2 Concept of µ-tools.

We have just defined the concept of µ-tool like which can bring a support to realization of an elementary and specific task entering in a well defined activity (table 1).

Table 1: Adapted levels of co-operative information systems according to the Activity Theory [24, 13, 15].

| *Level* | *Activity Theory* | *Human centered* | *System centered* |
|---|---|---|---|
| Macro | Activity | Motivation | Co-operative system |
| **Micro** | **Action/Task** | **Goal** | **Micro-tool** |
| Nano | Operation | Conditions | Functionality |

The product of a task can be an intermediate object of design. Development of µ-tools corresponds to an oriented step activity. We present below the various principles which lead to their data-processing structuring:
- Use of µ-tool is individual or collective; tasks can be structured in action plan. Then, it is necessary to specify the conditions of use of a µ-tool (life cycle of generated objects, share conditions...).
- Interaction between actor and µ-tool relates mainly to data acquisition (objects of activity), their relations, their accesses and their management, with graphic tools for example.
- Identification and description of µ-tool being the result of a collective and multi-field work; realization of models is recommended to facilitate exchange of ideas.
- µ-tools are developed according to a procedure of software quality which we defined in our laboratory.

µ-tools make it possible to divide software into modules adapted to achievement of simple tasks. Several tools in association can fulfill more complex functions. We defined a complete process of development of µ-tools for all actors cooperate (Figure 3). This process begins with the analysis of activity and leads on the corresponding software products and on the delivery of seven documents constituting the memory of their designs. Three great phases structure this process which we called *IDI* (**I**dentification, **D**esign, and **I**ntegration):
- *Phase of identification* of µ-tools, referring on the higher levels of engineering system and requiring a large collaboration of all actors, consists: 1) in identifying among the tasks which make a specific activity, those which could be instrumented, 2) then, to specify them.
- *Phase of design* aims, according to an incremental approach which facilitates permanent dialogue between all actors of development process, to define the architecture of µ-tool and these components, to develop them (UML/Java), to test them, and finally to validate the user interface.
- *Phase of integration* into *PLACID*, agent platform that is connected to an ORB (Object Broker Request, CORBA in fact) for exchanges management and information sharing.

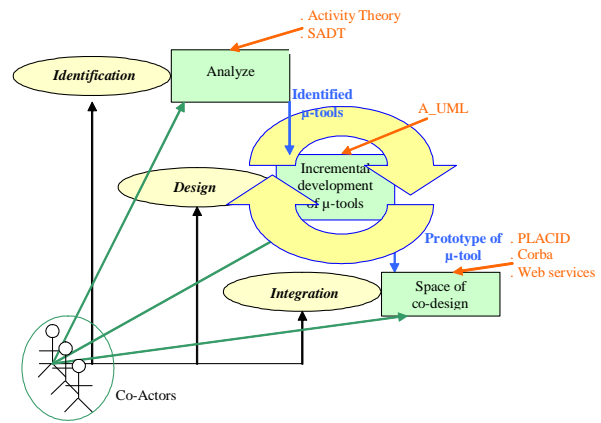

Fig. 3 Development process of µ-tool.

## 3. Agents for Co-operative Activities

Multi-agent systems have been proposed as a new approach for distributed artificial intelligence [27]. The main characteristics of agents (autonomy, distribution, adaptability, flexibility, cooperation and communication [29] permit, on the one hand, to manage efficiently the heterogeneous, autonomous and distributed solutions, and on the other hand, to facilitate exchanges of information and the sharing of resources between solutions (communication and cooperation). The idea of using the paradigm agent to design complex, interactive systems, either distributed or cooperative is very attractive. Indeed, very early, researchers found in the set of characteristics of agents, the means to design efficiently some cooperative information systems [26, 25, 3]. Other researchers, as Jennings [12], justified the adequacy of the approach agent for distributed system modeling and design (*adequacy hypothesis*).

MAS make it possible to distribute agents (processes) which are communicating entities, autonomous, reactive and having competences [22, 7]. To design MAS according to these criteria, it is necessary that each agent owns the three following properties: independence, communication and intelligence (expertise, skills, know-how). We also must define the architecture of agents (cognitive functions, interactions), and structure the knowledge necessary for their various activities. These properties correspond to those definite as well for an assistance platform to the collaborative design, as for the applications which it supports (all the more when they are co-operative µ-tools).

### 3.1 Agent Modeling

Agents being heterogeneous entities with various modes of interactions and complex behaviours, it is necessary to define their type of organization, and their capacity of evolution. Many definitions of paradigm agent have been proposed, one of the most consensual was made by Wooldridge. According to Wooldridge [28] an agent is an encapsulated computer system that is situated in some environment and that is capable of flexible, autonomous action in that environment in order to meet its design objectives. Furthermore, a software agent is rational, finalized and co-operative [29].

The autonomy of an agent is technically implemented by: 1) an independent process, 2) an individual memory (knowledge and data), and 3) ability to interact with other agents (perception or reception, emission or action).

The systemic model of MAS is defined as follows: the agents of MAS evolve in an environment and interact with each other, respecting the roles assigned to them in an organization. Then MAS are described by the following quadruplet (1):

$MAS = <Agt, Int, Ro, Co>$ (1)

Where *Agt* is the set of agents, *Int* is the set of interactions defined for these MAS, *Ro* is the set of roles to be played by agents and *Co* is the organization of agents into communities, when they are defined [18].

Many structures of agents known as "cognitive" are inspired by the cycle *<perceive, decide, act>*. However, our agent model [8, 9] is rather inspired by Rasmussen's three-level operator [19]: 1) reflex-based behaviour, 2) rule-based behaviour, and 3) knowledge-based behaviour with interpretation, decision and plan (Figure 4). We interpreted this model as a model of process for agents [17, 18]. The latter are both cognitive and reactive. Moreover, they have behaviours adapted to the tasks they perform. We added one level at this scale to include behaviour based on a system of agents. We call an actor (or collective agent) a system of cooperative agents in which the behaviour is defined by collective decision tasks and collective coordination tasks [17].

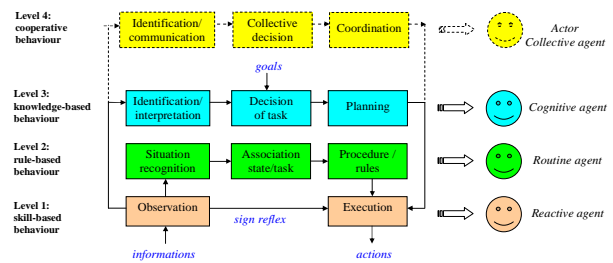

Fig. 4 Variability behaviors of agents, based on Rasmussen's model.

Thus, these agents (whose functional architecture and formalization in the form of UML class diagram are proposed in Figures 5.a and 5.b) may perform reflex actions, routine actions, or creative and cooperative actions in new situations. An agent that we consider in this paper, corresponds to the second level of Rasmussen's scale (Figure 4), and is described by the following quadruplet (2):

$Agent = <O, D, A, KB>$ (2)

Where *O* is the observation function of an agent; *D* is its decision function to interpret the observed events; *A* is its function of managing actions; and *KB* is the knowledge contained in its memory, among which are the decision rules and fuzzy values of the domain (the acquaintances and/or networks of affinities between agents), along with dynamic knowledge (observed events, internal states, etc.). The management of resources is provided by a set of managers $M = (Mm, Ma, Mk)$, where *Mm* is the message manager, *Ma* is the action manager and *Mk* is the knowledge-base manager (Figure 5a).

The decision rules of an agent (Rule), gathered in its knowledge base, are described by the following triplet (3):

$Rule = <E, C, Act>$ (3)

Where *E* is the set of events, *C* is the set of conditions and *Act* is the set of actions.

Actions of each agent are controlled by a manager *Ma* which memorizes tasks in the form (4):

$$Task = <Act, S, Re> \quad (4)$$

Where a task (*Task*) is characterized by a set of associated actions (*Act*), a set of states (*S*) and a set of results (*Re*).

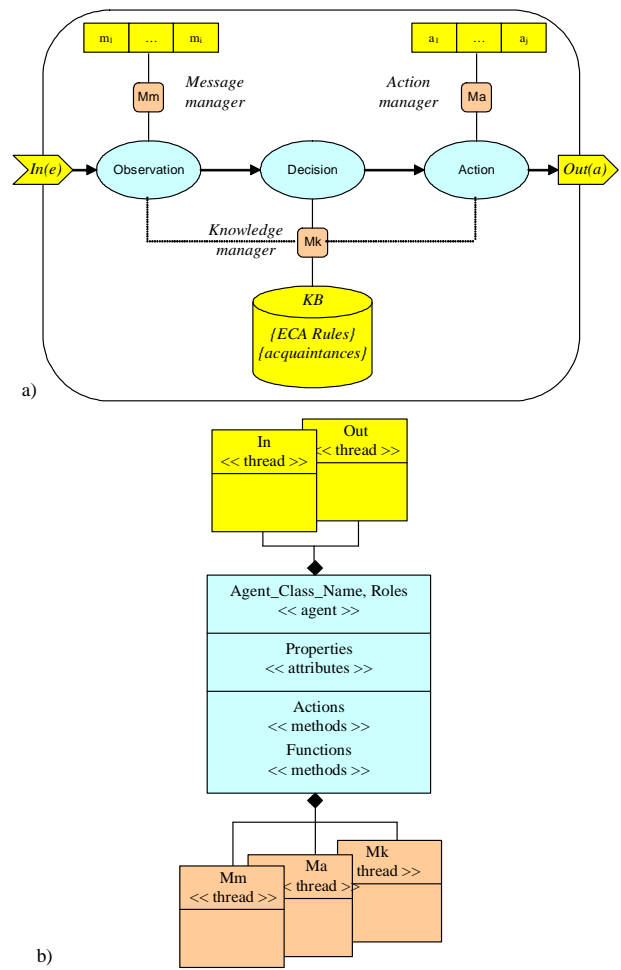

Fig. 5 Model of agent: a) functional architecture, and b) UML structure.

### 3.2 Communicating Agents

Communication between agents, following the HCI model, is characterized by: 1) a communication mode (sharing information or sending messages); 2) a common language; and 3) a communication protocol. To communicate with each other, agents exchange messages in a similar syntax KQML [14], an interaction language based on the concept of speech acts. These exchanges are controlled by a communication protocol in which a response is required for some speech acts (ask/answer, inform/confirm, etc.).

To design MAS, we developed a generic set of communication acts [17]. The agents perform five speech acts: inform, diffuse, ask, reply and confirm. The basic elements of this language (variables and primitives of the language) are listed in the following table (Table 2). These five speech acts are sufficient to enable agents to perceive the intention associated with the propositional content of a message. A communication act between two agents (CA) is then defined by the quintuplet (5):

$$CA = <l, x_e, x_r, t, m> \quad (5)$$

Where $l$ is a speech act denoted by a performative verb, $x_e$ is the source of communication, $x_r$ is the receiver, $t$ is the type of message and $m$ is the message itself, which can be an assertion, a question, a response, etc.

Table 2: Elements of interaction language

| Language | Meaning |
| --- | --- |
| x, e, a, m, t | respectively are agent, event, action, message and type of message |
| inform($x_e$, $x_r$, t, m) | $x_e$ sends to $x_r$ the message m of type t |
| diffuse($x_e$, $x_i$, t, m) | $x_e$ sends to the list $x_i$ the message m of type t |
| ask($x_e$, $x_r$, t, m) | $x_e$ asks $x_r$ the request m of type t |
| answer($x_e$, $x_r$, t, m) | $x_e$ answers $x_r$ the message m of type t |
| confirm($x_e$, $x_r$, t, m) | $x_e$ confirms to $x_r$ that it agree with the message m of type t |

### 3.3 Co-operating Agents

The systems of co-operative work consist of distributed, heterogeneous and autonomous components. Then, MAS are well adapted. The potential contribution of paradigm agents concerns:

- A more natural interactivity (methods, presentation).
- The management of repetitive actions and the delegation of tasks without interest for the user.
- The decision-making by the comprehension of the context of use (relevance).
- The personalization of information (preferences, goals and capacities of the user).

The individual and co-operative behaviors of the agents are varied: initializations, planning of actions, emission and reception of documents, information or document retrieval, supervision of procedures, etc. Each one of these services corresponds to a competence.

### 3.4 Process of Agentification

The agents are entities having competences which enable them to play one or more roles in an organization. They are grouped within MAS organized according to a hierarchical structure (three types of agents: specialists, mediators and supervisors). For the specification and the conceptualization of MAS [2] we retain proposals made in the definition of the language A_UML [16, 1], like our own methodological proposals [8] (see Figure 6):

1) To design the use case diagram (services provided by the system), and for each identified use carried out 3 following phases;
2) To design the classes diagram connecting the agents concerned with the use (we can also use the collaboration diagram);
3) To define behavior of each agent with a states diagram or an activities diagram;
4) On the basis of scenarios of use, to design the sequence diagrams which specify the exchanges of messages between agents (and their scheduling).

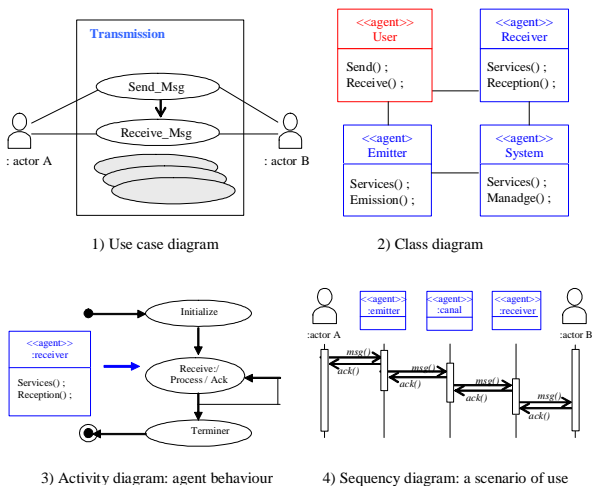

Fig. 6 Methodology followed for the illustration of §5.1.

## 4. Description of *PLACID*

### 4.1 Project Objectives

The development of *PLACID* is defined in the 2 research orientations of our team (first, point of view on the product, and secondly, co-operation in design), in order to facilitate the use of design µ-tools by a team of close or distant designers, within the activity of distributed design, structured in modules (functional, structural, manufacture, maintenance). *PLACID* offers a set of services for use of an environment of virtual co-design (shared objects, services of tasks management and communications, and tools of decision-making aid).

Figure 7 presents the modular architecture of the platform. The 5 layers defined allow multi-platforms uses and facilitates its evolution:
– Layer 1: level of user interface, with a context of many interactions, multi-user and multimode.
– Layer 2: level of workspace organized according to the context of an individual or co-operative activity.
– Layer 3: level of management of Co-design assistance tools (µ-tools and other design tools).
– Layer 4: level of management of co-operative work, allowing to control and to carry out co-operative processes.
– Layer 5: level of operating system and management of low level communications.

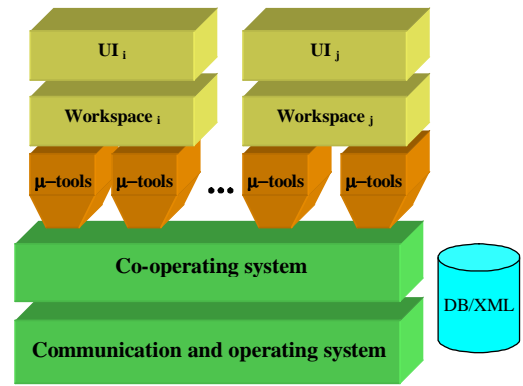

Fig. 7 Modular architecture of *PLACID* platform.

For data management we retained an architecture adapted to co-operative information systems [25]: a federating DBMS of multi-bases. This makes it possible to run total or local applications.

### 4.2 Architecture of *PLACID*

The various possible configurations of *PLACID* platform as well allow specific use of a design µ-tool, as constitution of "virtual desks of Co-design". This is facilitated by the agent-based design of *PLACID*. The first layer of platform (see figure 8) is connected to CORBA (Common Object Request Broker Architecture) in charge of management of exchanges and shared information. This layer of agents is composed of two groups: "mediators" agents, in direct interaction with users and µ-tools, and "tools" agents (executants) that have essential skills of co-operation.

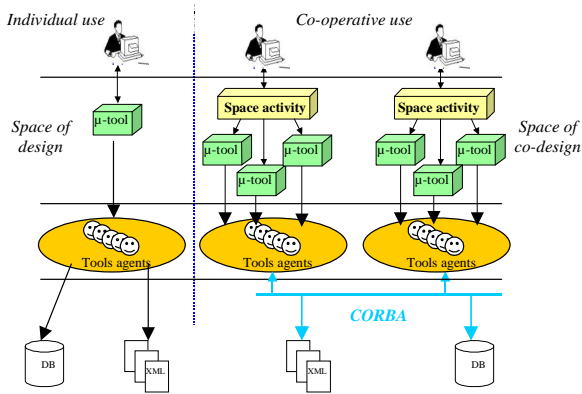

Fig. 8 Various configurations to use the platform.

## 5. Illustration of µ-Tools Integration: the *Papoticiel*

To test *PLACID* platform according to points of view presented above, we decided to develop 2 types of µ-tools: the first correspond to an electronic meeting for distant and collaborative use (*Papoticiel,* a chat tool), and the second allow carrying out a functional analysis in design activity. Following we present the first set of µ-tools.

An application of electronic meeting, in addition to the communication aspects, connects applications of co-operation as varied as: management of a group, maintenance of a diary, management of a group memory through the filing of discussions in meetings, edition of minutes of meeting. We describe below, some elements of *Papoticiel* µ-tools design, according to the methodology presented in figure 6:
- Use case diagram (see Figure 9) defines context of use of *Papoticiel*. This application can be started on initiative of a group member (initiator) or by the software agent <diary>. SADT diagram (see Figure 10) can also design to represent tasks model of use of chat (for example: to open the chat, to check presences, to discuss, to close the chat).
- Class diagram of electronic meeting to define the structure of agents, and the interrelationships between agents (this one can be supplemented by a collaboration diagram).
- Finally, sequence diagram or collaboration diagram. The following figure (Figure 11) illustrates the scenario "begin-end" of these µ-tools, started by a participant of meeting (initiator).

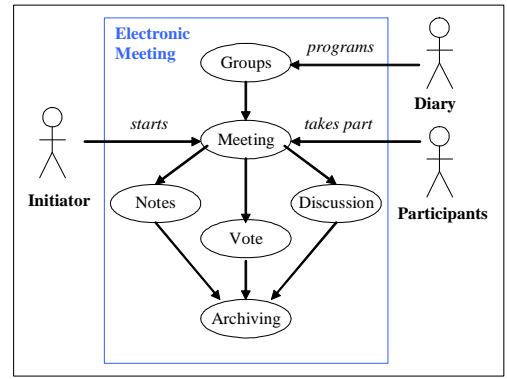

Fig. 9 Use-case diagram of *Papoticiel*.

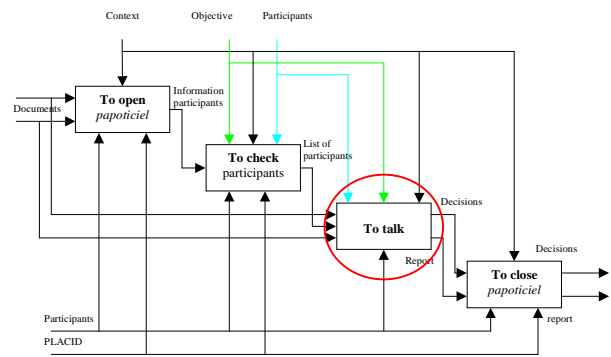

Fig. 10 SADT diagram of the task «*Using Papoticiel*».

To design µ-tools of electronic meeting, illustrated by the preceding figures, we needed 7 types of agents which are described in table 3. Agent structure of *Papoticiel* above *PLACID* platform is presented in figure 12. *Papoticiel*, in addition to agents already available on *PLACID* platform, requires the deployment of 2 other agents (<customerCorba> agent and <serverCorba> agent) to manage co-operation as well as multi-user context.

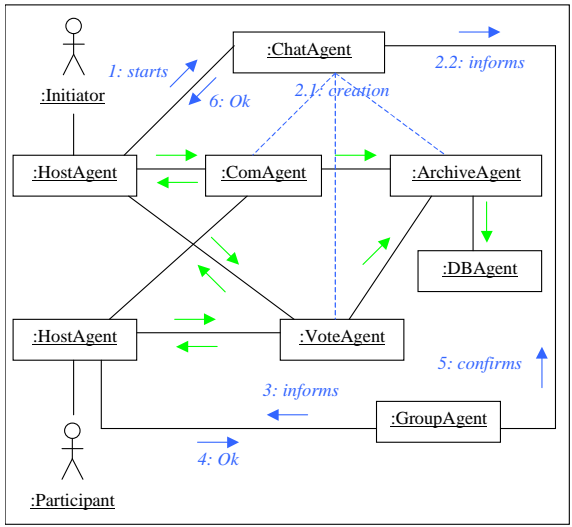

Fig. 11 Collaboration diagram illustrating scenario "begin-end".

Table 3: Competences of the various agents of *Papoticiel*

| Agents | Skills |
|---|---|
| <User> | • Authentication and user identification<br>• Access management towards other agents<br>• Access management to other resources/services<br>• Transmission of messages |
| <Com> | • Identification of messages' writers<br>• Messages queues management<br>• Sending of messages to agents<br>• Sending of messages to participants |
| <Papoticiel> | • Access management to application<br>• Queues management to avoid conflicts<br>• Sending of messages to agents |
| <DB> | • Data update<br>• Sending of messages<br>• Sending of results (during a later consultation)<br>• Saving of messages and results |
| <Group> | • Management of various working groups<br>• User identification<br>• Connection with other services (or agents) |
| <Archive> | • User identification<br>• Reception of messages<br>• Filing of messages<br>• Filing outcome of votes |
| <Vote> | • Activation of beginning and end of votes<br>• Sending the voting results<br>• Connection with agents |

Integration of *Papoticiel* into *PLACID* platform is carried out in 2 phases: the first corresponds to the agentification of μ-tools composing *Papoticiel*, the second consists in defining then including <Corba> agent on the communication level of *Papoticiel*. To finish this description, we present the user interface of a prototype of workspace for functional analysis activity [17] (see figure 14). This figure shows the 3 μ-tools of the *Papoticiel* which are integrated into *PLACID* (Papoticiel/chat tool, Agenda/diary tool, Vote/voting tool).

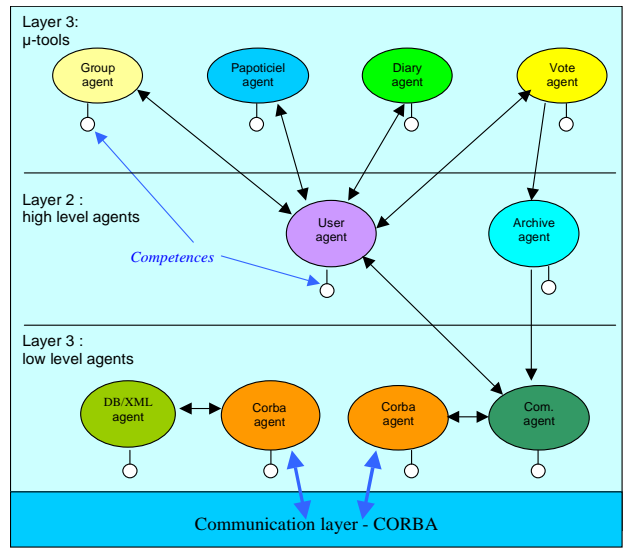

Fig. 12 Agent structure of *Papoticiel* on *PLACID* platform.

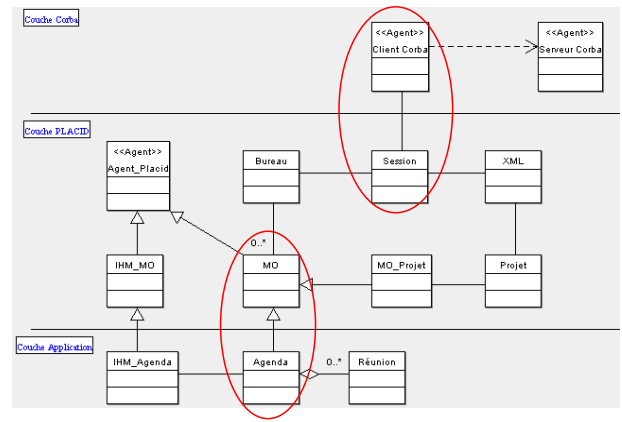

Fig. 13 Integration of the μ-tool "Agenda/Diary" into *PLACID* platform.

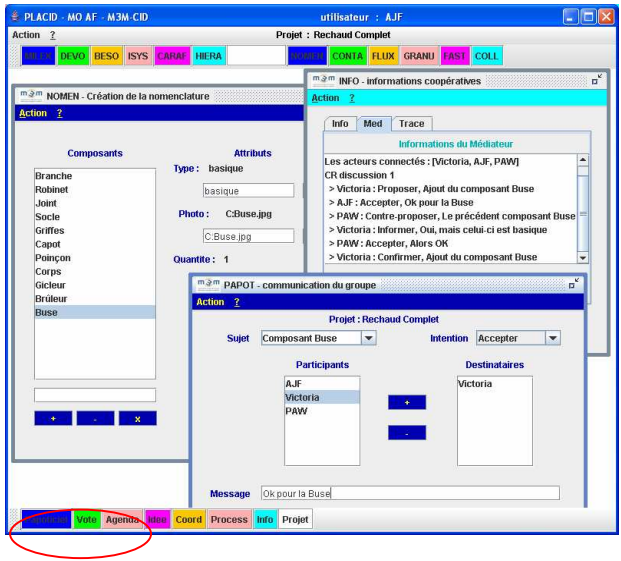

Fig. 14 Functional Analysis Workspace and µ-tools of *Papoticiel*.

## 6. Conclusions

A first study on prospects for co-operative work and agents technologies had provided us with a set of promising concepts which were used for the definition of *PLACID* platform:
- About *CSCW*: definition of project of distributed co-design, supporting services of shared resources, of group coordination, of electronic meeting, of group decisions, report of working session and of group memories.
- About *MAS*: distribution of co-operative activities, distribution of components of design assistance; decision-making aid in the process of distributed design, management of actions (repetitive or implicit tasks within a cooperative activity), design of human-computer interfaces.

Following this study and a first phase of specification of *PLACID*, we produced a prototype of the platform. This one was tested by the use of 2 sets of µ-tools (until other µ-tools are developed then integrated into the platform):
- The first makes it possible to use and to coordinate an elementary electronic meeting coupled to a tool for assistance to groups' management; so we could test the synchronization and the distant exchange.
- The second corresponds to 6 modules deployed to carry out an *External Functional Analysis* during product design activity; so we could test the chaining process and the use of a database.

Since, we launched the development of three other sets of µ-tools for co-design: a first for assistance to the deployment of *TRIZ* methodology (Teoria Reschenia Izobretateliskih Zadaci, theory of resolution of the innovating problems), a second for assistance to the *Performance Evaluation* in engineering of manufacturing systems, and a third for the *Technical Functional Analysis*.

Our current work concerns on identification and definition of generic *CMT* (Co-operative Micro-Tools), and generalization of µ-tools-based assistance to any type of co-operative activities, in particular in the field of software engineering [17, 18].

### Acknowledgments

We gratefully acknowledge the PRéCI (Regional Center of Design and Innovation – Franche-Comté, France) for supporting our research topic "Design of Innovative Micro-Tools", as part of its plan of action "Anticipating the needs of SMEs in terms of design and innovation".

**Alain-Jérôme Fougères** is Computer Engineer and PhD in Artificial Intelligence from the University of Technology of Compiègne. He is currently a member of the Laboratory of Mecatronics3M (M3M) at the University of Technology of Belfort – Montbéliard (UTBM), where he conducts his research on cooperation in design. His areas of interests and scientific contributions concern the natural language processing, the knowledge representation, the design of multi-agent systems, in particular architecture, interactions, communication and co-operation problems. In recent years, his research has focused on the context of co-operative work (mediation of cooperation and context sharing), mainly in the field of mechanical systems co-design.